\documentclass[aps,prl,twocolumn,superscriptaddress,floatfix,longbibliography,10pt]{revtex4-2}

\usepackage{graphicx}
\usepackage{hyperref}
\usepackage{float} 
\usepackage{xcolor}
\usepackage{amsmath}
\usepackage{amssymb} 
\usepackage[T1]{fontenc}
\usepackage[english]{babel}
\usepackage{orcidlink}

\makeatletter
\let\ORIbbl@fixname\bbl@fixname
\def\bbl@fixname#1{%
  \@ifundefined{languagealias@\expandafter\string#1}
    {\ORIbbl@fixname#1}
    {\edef\languagename{\@nameuse{languagealias@#1}}}%
}
\newcommand{\definelanguagealias}[2]{%
  \@namedef{languagealias@#1}{#2}%
}
\makeatother
\definelanguagealias{en}{english}

\begin{document}


\title{New Limit on Axion-Like Dark Matter using Cold Neutrons}


\author{Ivo~Schulthess\orcidlink{0000-0002-5621-2462}}
\email[corresponding author: ]{ivo.schulthess@lhep.unibe.ch}
\affiliation{Laboratory for High Energy Physics and Albert Einstein Center for Fundamental Physics, University of Bern, 3012 Bern, Switzerland}

\author{Estelle~Chanel}
\altaffiliation[current position: ]{Institut Laue-Langevin, CS 20156, 38042 Grenoble Cedex 9, France}
\affiliation{Laboratory for High Energy Physics and Albert Einstein Center for Fundamental Physics, University of Bern, 3012 Bern, Switzerland}

\author{Anastasio~Fratangelo\orcidlink{0000-0001-9964-601X}}
\affiliation{Laboratory for High Energy Physics and Albert Einstein Center for Fundamental Physics, University of Bern, 3012 Bern, Switzerland}

\author{Alexander~Gottstein\orcidlink{0000-0002-9655-2728}}
\affiliation{Laboratory for High Energy Physics and Albert Einstein Center for Fundamental Physics, University of Bern, 3012 Bern, Switzerland}

\author{Andreas~Gsponer\orcidlink{0000-0002-5012-7371}}
\affiliation{Laboratory for High Energy Physics and Albert Einstein Center for Fundamental Physics, University of Bern, 3012 Bern, Switzerland}

\author{Zachary~Hodge\orcidlink{0000-0002-7004-168X}}
\altaffiliation[current position: ]{University of Wisconsin, Madison, Wisconsin, USA}
\affiliation{Laboratory for High Energy Physics and Albert Einstein Center for Fundamental Physics, University of Bern, 3012 Bern, Switzerland}

\author{Ciro~Pistillo\orcidlink{0000-0001-8131-9440}}
\affiliation{Laboratory for High Energy Physics and Albert Einstein Center for Fundamental Physics, University of Bern, 3012 Bern, Switzerland}

\author{Dieter~Ries\orcidlink{0000-0003-1663-6989}}
\altaffiliation[current position: ]{Department of Chemistry - TRIGA site, Johannes Gutenberg University Mainz, 55128 Mainz, Germany}
\affiliation{Laboratory for High Energy Physics and Albert Einstein Center for Fundamental Physics, University of Bern, 3012 Bern, Switzerland}

\author{Torsten~Soldner\orcidlink{0000-0003-2468-5182}}
\affiliation{Institut Laue-Langevin, CS 20156, 38042 Grenoble Cedex 9, France}

\author{Jacob~Thorne\orcidlink{0000-0002-3905-5549}}
\affiliation{Laboratory for High Energy Physics and Albert Einstein Center for Fundamental Physics, University of Bern, 3012 Bern, Switzerland}

\author{Florian~M.~Piegsa\orcidlink{0000-0002-4393-1054}}
\email[corresponding author: ]{florian.piegsa@lhep.unibe.ch}
\affiliation{Laboratory for High Energy Physics and Albert Einstein Center for Fundamental Physics, University of Bern, 3012 Bern, Switzerland}


\date{\today}

\begin{abstract}
We report on a search for dark matter axion-like particles (ALPs) using a Ramsey-type apparatus for cold neutrons. A hypothetical ALP-gluon-coupling would manifest in a neutron electric dipole moment signal oscillating in time. 
Twenty-four hours of data have been analyzed in a frequency range from 23~$\mu$Hz to 1~kHz, and no significant oscillating signal has been found. 
The usage of present dark-matter models allows to constrain the coupling of ALPs to gluons in the mass range from $10^{-19}$ to $4 \times 10^{-12}$~eV. The best limit of $C_G$~/~$f_a m_a = 2.7 \times 10^{13}$~GeV$^{-2}$ (95\%~C.L.) is reached in the mass range from $2 \times 10^{-17}$ to $2 \times 10^{-14}$~eV.
\end{abstract}


\maketitle



Dark matter makes up roughly 27\% of our universe's total mass-energy content~\cite{planck_collaboration_planck_2020}. 
So far, no dark-matter model has been experimentally verified, but promising candidates remain the axion and a more general class of axion-like particles (ALPs) with relaxed property constraints. 
The axion was initially suggested to solve the strong CP problem of quantum chromodynamics (QCD)~\cite{peccei_cp_1977,peccei_constraints_1977,weinberg_new_1978,wilczek_problem_1978}. The solution is an additional U(1) symmetry to the Standard Model of particle physics. The spontaneous and explicit breaking of this symmetry results in a massive but ultra-light \mbox{spin-0} particle, the axion. Since this pseudo-scalar particle must satisfy the Klein-Gordon equation, it results in an oscillating field that could explain the dark-matter content in our universe. 
Most experiments, such as CAST, IAXO, or ADMX search for the axion via its coupling to photons~\cite{dafni_axion_2015,braine_extended_2020}. 
Various models suggest interactions with other particles such as standard model fermions (DFSZ models~\cite{dine_simple_1981, zhitnitskii_possible_1980}) or a new exotic heavy quark (KSVZ models~\cite{kim_weak-interaction_1979, shifman_can_1980}). A summary of recent axion theories and experiments can be found in the PDG review~\cite{zyla_review_2020}. 
The coupling of axions and ALPs to gluons is a common feature in theoretical models~\cite{g_g_raffelt_stars_1996, hook_solving_2018, di_luzio_dark_2021, di_luzio_even_2021}. 
One consequence of this coupling is that an oscillating ALP field induces an equally oscillating electric dipole moment (EDM) of the neutron~\cite{pospelov_theta_2000}
\begin{equation}\label{eq:axionGluonEDM}
    d_n^{a}(t) \approx +2.4 \times 10^{-16} \, e \, \mathrm{cm} \cdot \frac{C_G}{f_a} a_0 \cos{(m_a t)}  \ , 
\end{equation}
where $C_G$ is a model-dependent parameter, $f_a$ the ALP decay constant, $a_0$ its oscillation amplitude, $m_a$ its mass, and $e$ the elementary charge.
The parameter space of ALPs is defined by their mass and the coupling $C_G/f_a$. It is restricted by various astrophysical and cosmological constraints, as well as scrutinized in three recent laboratory experiments. 
The CASPEr experiment is dedicated to searching for axion signals using nuclear magnetic resonance techniques~\cite{budker_proposal_2014,aybas_search_2021}. 
Two other experiments search for a permanent EDM of the electron, using trapped molecular ions, and the neutron, using ultra-cold neutrons in a storage experiment. 
Both experiments found no significant oscillating signal from the nHz-region up to 0.4~Hz~\cite{abel_search_2017, roussy_experimental_2021}. 
Here, we present the results of the Beam EDM experiment, which employs a continuous cold neutron beam with intrinsic sub-ms time resolution~\cite{piegsa_new_2013, chanel_pulsed_2019}. 
Thus, the accessible frequency range is extended to 1~kHz. This allowed us to extend the probed ALP-mass range by more than three orders of magnitude. 
Since no significant signal was found, a new constraint on the possible existence of such ultra-light particles has been deduced.\\


We use Ramsey's method of separated oscillatory fields applied to neutrons to search for an axion-like dark-matter signal~\cite{ramsey_new_1949,ramsey_molecular_1950}. In this technique, neutrons act as a spin clock at their Larmor precession frequency in a magnetic field $B_0$, which allows to precisely detect magnetic or pseudo-magnetic field changes. 
The measured quantity is the phase that a neutron spin acquires due to its coupling to a magnetic field change $\Delta B (t)$ and an electric field $E$ 
\begin{equation}\label{eq:precession}
    \varphi = \displaystyle \int_0^{T_\text{int}} \left( \gamma_n \Delta B (t) + \frac{2 d_n(t)}{\hbar} E \right) \, \mathrm{d}t \ ,
\end{equation}
where $\gamma_n$ is the gyromagnetic ratio of the neutron, $d_n (t)$ its electric dipole moment, $\hbar$ the reduced Planck constant, and $T_\text{int}$ the interaction time which depends on the neutron velocity. 
\begin{figure}
    \centering
    \includegraphics[width=0.48\textwidth]{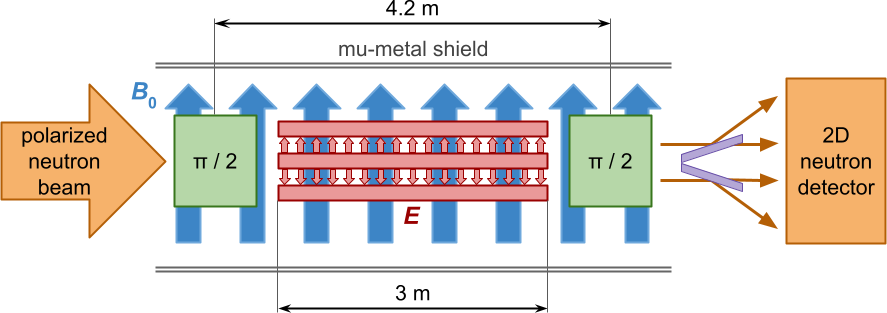}
    \caption{\label{fig:setupSchematic}(Color online) Schematic of the experimental setup where a neutron beam enters from the left, polarized along the $B_0$-field. It shows the 6~m-long mu-metal shield around the interaction region and the two 40~cm-long RF spin-flip coils for the $\pi/2$-flips in green. The electrodes and the electric field direction are shown in red and the magnetic field direction is indicated in blue. The spin analyzer (purple) reflects one spin state and transmits the other. The neutrons are detected using a 2D pixel detector with a sensitive area of $10 \times 10$~cm$^2$ with $16 \times 16$~pixels. The vacuum beam pipe surrounding the electrodes is not shown. }
\end{figure}
Figure~\ref{fig:setupSchematic} shows a schematic of the experimental setup installed at the cold neutron beam facility PF1b at the Institut Laue-Langevin in Grenoble, France~\cite{abele_characterization_2006}. 
A beam of polarized cold neutrons with a Maxwell-Boltzmann-like velocity distribution, peaking at about 1000~m/s, enters a constant and homogeneous vertical magnetic field $B_0 = 220$~$\mu$T. The field is adjusted and stabilized to the sub-nT level using fluxgate sensors and a 3D~coil system. 
A two-layer passive magnetic mu-metal shield surrounds the experimental setup. 
Two radio frequency (RF) spin-flip coils, one before and one after the interaction region, induce resonant $\pi$/2-flips of the neutron spin. 
The interaction region with a length of 3~m is inside a vacuum beam pipe between the spin-flip coils. 
It consists of three sets of one-meter-long electrode stacks with a high-voltage electrode in the center and two ground electrodes on top and bottom. 
The electrode separation is 1~cm. 
This setup allows for two neutron beams passing between the electrodes, simultaneously sensing the electric field direction parallel and anti-parallel to the magnetic field. This double beam arrangement provides the possibility to compensate for global field drifts and common-mode noise. 
Downstream of the setup, a neutron spin analyzer spatially separates the two spin states of each beam before they are counted in a 2D~neutron pixel detector~\cite{klein_cascade_2011}.
The neutron rate integrated over the entire sensitive area of the detector was approximately $10^7$~s$^{-1}$. The statistical counting error of the detector was calibrated. It was found that the Poisson error is overestimating the measured standard deviation by approximately 10\% for the given neutron rate and settings of data acquisition due to event pile-up.
The potential of the high-voltage electrode was set to $\pm 35$~kV, and the resulting electric field was directly measured with neutrons using the relativistic $\vec{v} \times \vec{E}$-effect~\cite{dress_search_1977}.
The measured electric field amplitude agrees with the nominal value within 4\%. The reason for the small deviation is the slight vertical displacement of the central high-voltage electrode due to gravity.
With this apparatus, the oscillating neutron EDM caused by a hypothetical axion field would manifest in an oscillation of the population in each spin state. This would lead to a corresponding oscillation of the neutron asymmetry for each beam, defined as  
\begin{equation}\label{eq:neutronAsymmetry}
    \mathcal{A} = \frac{N_{\uparrow} - N_{\downarrow}}{N_{\uparrow} + N_{\downarrow}} \ , 
\end{equation}
where $N_{\uparrow}$ and $N_{\downarrow}$ are the neutron counts in the spin up and down state, respectively.
To be most sensitive to changes in the asymmetry, the frequency and relative phase of the RF spin-flip signals are adjusted such that for each beam $\mathcal{A} \approx 0$, i.e.,\ $N_{\uparrow} \approx N_{\downarrow}$. This corresponds to the point of steepest slope in a Ramsey resonance pattern. \\

\begin{figure}
    \centering
    \includegraphics[width=0.48\textwidth]{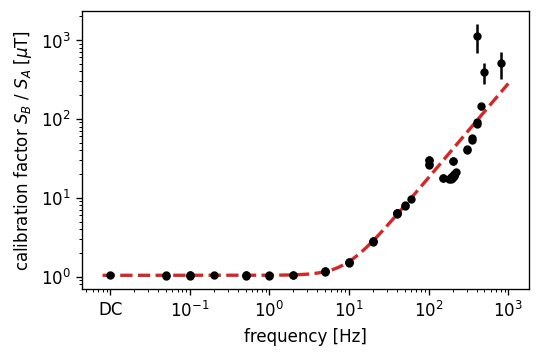}
   \caption{\label{fig:calibration}(Color online) Calibration factor $S_B / S_A$ as a function of frequency. The measured data are shown as dots, whereas the red dashed line is a least-squares fit of a Butterworth-filter function~\cite{stephen_butterworth_theory_1930} which is used for the data analysis. 
   For instance, typical neutron asymmetry signals of the order $10^{-5}$ correspond to a pseudo-magnetic field of $14$~pT for frequencies smaller than 5~Hz using Eq.~(\ref{eq:magneticFieldFromAsymmetry}). }
\end{figure}

To connect the signal amplitude of the neutron asymmetry of Eq.~(\ref{eq:neutronAsymmetry}) to the ALP-gluon-coupling in Eq.~(\ref{eq:axionGluonEDM}) multiple calibration measurements were conducted. 
In these measurements, we created artificial signals by applying homogeneous sinusoidally oscillating magnetic fields of various frequencies and amplitudes $B_a$ parallel to $B_0$ through the entire setup using an additional rectangular Helmholtz-type coil. 
Note, such a field can be interpreted as a corresponding false EDM signal using Eq.~(\ref{eq:precession})
\begin{equation}\label{eq:faxionCorrelation}
   d_n = \frac{\hbar \gamma_n B_a}{2 E}.
\end{equation}
First, we conducted an offline calibration measurement where we correlated the magnetic field amplitude $B_a$ to the applied oscillating electric current in the auxiliary coil. 
The field was determined at 47~positions over a distance of 5.3~m along the neutron beam path with five fluxgates mounted in a cross-shaped arrangement on a magnetic field mapper. The magnetic field was recorded with a sampling rate of 10~kHz for two seconds at each position and a sinusoidal function was fitted to the data. The amplitude was averaged over the interaction region and all five fluxgates. The calibration parameter was measured to $S_B = (12.13 \pm 0.02)$~$\mu$T~A$^{-1}$. \\
A second calibration measurement was performed with neutrons to correlate the amplitude of the oscillating neutron asymmetry in Eq.~(\ref{eq:neutronAsymmetry}) to the same coil currents applied in the first calibration measurement. Here, we acquired the neutron asymmetry for 60~seconds at a sampling rate of 4~kHz and performed again a sinusoidal fit to the data. This resulted in a value of $S_A = (11.5 \pm 0.5)$~A$^{-1}$ for frequencies below 5~Hz.
Together, the two calibration measurements are used to translate the amplitude of an oscillating neutron asymmetry into a corresponding (pseudo-)magnetic field amplitude via
\begin{equation}\label{eq:magneticFieldFromAsymmetry}
    B_a = \frac{4.2}{3}\frac{S_B}{S_A} \, \mathcal{A} \ .
\end{equation}
The factor of 4.2/3 comes from the fact that the magnetic and the electric interaction length are different as shown in Fig.~\ref{fig:setupSchematic}. 
The resulting calibration curve as a function of frequency is presented in Fig.~\ref{fig:calibration}. 
The value is constant for low frequencies up to approximately 5~Hz. 
The primary reason for its subsequent rise is the frequency-dependent RF shielding of the aluminum parts of the setup, i.e.,\ the construction frame, vacuum beam pipe, and electrodes. 
Another reason is an effect that depends on the neutron velocity: 
as shown in Eq.~(\ref{eq:precession}), the acquired neutron spin phase has to be integrated over the interaction time. 
In the case of an oscillating field, this integral becomes zero if the period of the oscillation matches the interaction time. This effect is suppressed for a beam with a broad velocity distribution but still results in a decrease in sensitivity at higher frequencies. Calculations, simulations, and further test measurements that are not included in this paper for brevity suggest that the actual decrease in sensitivity would be smaller for real ALP signals. 
However, since these effects cannot be simply decorrelated, we use the presented curve. This results in a conservative upper limit at high frequencies if no ALPs were found. \\

\begin{figure*}
    \centering
    \includegraphics[width=0.90\textwidth]{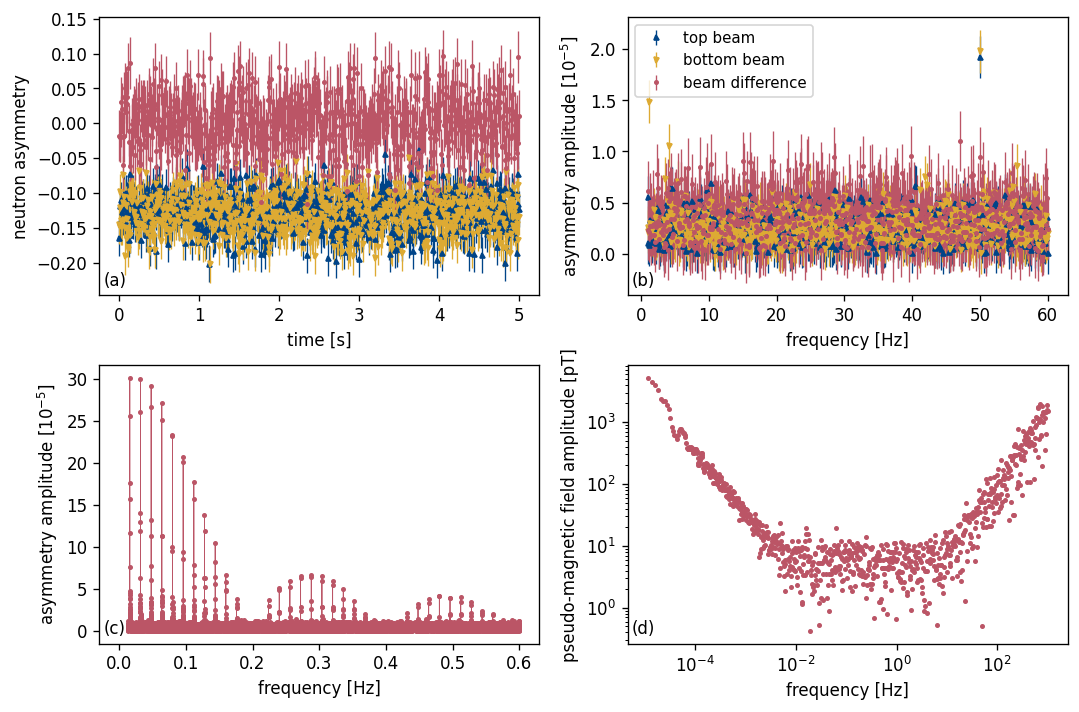}
    \caption{\label{fig:analysisSteps}(Color online) The data for the top beam (blue $\blacktriangle$), the bottom beam (yellow $\blacktriangledown$), and the difference between the two beams (red $\bullet$) are shown for various stages in the data processing. (a)~Measured neutron asymmetry for a time window of 5~seconds. (b)~Frequency spectrum between 1~Hz and 60~Hz. The highly significant signal at 50~Hz from the power line frequency are canceled out by analyzing the beam difference. (c)~Discrete peaks appear in the spectrum due to the data structure. (d)~After applying the calibration as shown in Fig.~\ref{fig:calibration}, the neutron asymmetry spectrum translates into a pseudo-magnetic field spectrum. Error bars were omitted for reasons of readability but are of order 3~pT in the most sensitive central range. Note that the plots of (b), (c), and (d) are based on the whole 12 hour data set and that only a fraction of the data points is shown for legibility in all sub-figures.  }
\end{figure*}

\begin{figure}
    \centering
    \includegraphics[width=0.48\textwidth]{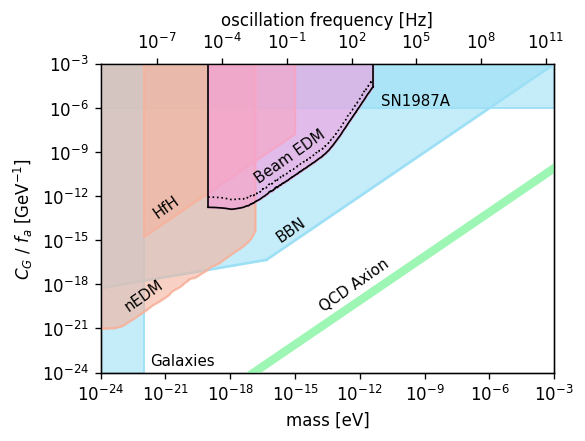}
    \caption{\label{fig:exclusionPlot}(Color online) Limits on the ALP-gluon-coupling are shown as a function of the mass or frequency. The shaded areas are exclusion regions from cosmology and astrophysical observations (blue: Galaxies~\cite{corasaniti_constraints_2017}, BBN~\cite{blum_constraining_2014,stadnik_can_2015}, SN1987A~\cite{raffelt_astrophysical_1990, graham_new_2013}) and laboratory experiments (orange: nEDM~\cite{abel_search_2017}, HfH~\cite{roussy_experimental_2021}). The black outlines with the pink area mark the exclusion region of this publication (labeled Beam EDM). The solid and dotted lines correspond to the deterministic and stochastic dark-matter models, respectively. The green line shows the canonical QCD axion. }
\end{figure}

We performed several continuous measurements of the neutron EDM for the dedicated ALP search with various duration and high-voltage polarities. 
The presented analysis uses a total of 24~hours of data, taken with a sampling rate of 4~kHz, i.e.,\ we obtained a value for the neutron asymmetry and, hence, the neutron EDM every 0.25~ms. The potential of the central high-voltage electrode was set to +35~kV. Hence, the electric field used for the evaluation is $E=2 \times 35$~kV/cm as all the analysis is done for the difference of the two beams. Data were taken on September 13/14, 2020 and are publicly available~\cite{piegsa_new_2020}. 
A 5~second-long subset of the data is presented in Fig.~\ref{fig:analysisSteps}a. 
The entire data is split into two halves of 12~hours each. Limits are based on the first half of the data, but an oscillating signal would only be considered significant if it appears in the spectral analysis of both sets at least on the 5-sigma level. 
We performed the spectral analysis on the neutron data using an adapted version of the generalized Lomb-Scargle algorithm~\cite{lomb_least-squares_1976, zechmeister_generalised_2009, press_fast_1989, vanderplas_understanding_2018}. 
The basic concept of the algorithm is to perform a $\chi^2$ minimization of the fit function $ f(t) = a \sin(\omega t) + b \cos(\omega t) + c $, where $a$, $b$, and $c$ are the parameters to be minimized for each frequency $\omega$. 
The signal amplitude $\sqrt{a^2 + b^2}$ is Rayleigh distributed assuming only white noise.\\
We subtracted the signals of both neutron beams, i.e.,\ opposite electric field directions, from each other. Figure~\ref{fig:analysisSteps}b shows how this eliminates the eminent 50~Hz signal coming from the power line frequency. The peak is visible in the spectra of both beams separately but not in the spectrum of the difference of the signals. 
The overall spectrum shows three groups of significant signals of different origins that are not ALPs.
The first group appears for frequencies below 10~mHz. They can be explained by long-term magnetic gradient field drifts due to temperature changes. They happen on the time scale of hours and result in a rise in signal amplitude. 
The second group is located in the frequency range between 10~mHz and 2~Hz. They are caused by the data structure itself and a sub-range is presented in Fig.~\ref{fig:analysisSteps}c. Our sequence of data taking is divided into runs of 62.5~s duration.
Each run consists of 57.5~s of measurement time and 5~s of downtime to save the data.
This time structure leads to peaks at the inverse run time of 16~mHz and higher orders. Additionally, the 5~s gap leads to an envelop hump structure with a period of 200~mHz. For frequencies higher than 2~Hz, these peaks are too small to be detected. 
The third group of significant signals has a statistical origin. Since the amplitudes of the signals follow a Rayleigh distribution, the pull/significance is also distributed accordingly. We found $113 \pm 11_{\text{stat}} \pm 9_{\text{sys}}$ and $132 \pm 11_{\text{stat}} \pm 13_{\text{sys}}$ events above the 5-sigma threshold in the first and second half of the data, respectively. The frequencies of the events of both sets do not coincide. The systematic error originates from the uncertainty of the detector count-error calibration. These values are slightly below the 162 statistically expected events for a data set with 43.6~million analyzed frequencies.
With the use of Eq.~(\ref{eq:magneticFieldFromAsymmetry}), the neutron asymmetry amplitude shown in Fig.~\ref{fig:analysisSteps}b and Fig.~\ref{fig:analysisSteps}c can be translated into the pseudo-magnetic field amplitude. The full spectrum with a reduced spectral resolution is shown in Fig.~\ref{fig:analysisSteps}d. The most sensitive region of a few pT is in the central flat region. \\
Besides having a significant amplitude over the background noise level, an actual oscillating EDM signal must disappear if no electric field is applied. This way, noise signals or signals from external sources can be further excluded. For this, we performed an additional measurement with no electric field applied. Moreover, the amplitude of a real signal must be identical for both electric field directions but must exhibit a phase-shift of $\pi$. \\
Overall, no significant signal was found at the same frequency in both partial data sets. Thus, an upper limit on the ALP-gluon-coupling can be derived. 
Using Eqs.~(\ref{eq:axionGluonEDM}) and (\ref{eq:faxionCorrelation}) as well as the calibration shown in Fig.~\ref{fig:calibration}, the coupling can be calculated with
\begin{equation}\label{eq:axionCoupling}
    \frac{C_G}{f_a} = \frac{\gamma_n \hbar B_a}{a_0 E \times 4.8 \times 10^{-16} \, e \, \text{cm}}. 
\end{equation}\\
The oscillation amplitude relates to the local dark-matter density via $a_0 = \sqrt{2\rho_{\text{\tiny{DM}}}} / m_a$, assuming all dark matter consists of ALPs. 
The coherence time of the dark-matter field $\tau_c$ is $10^6$ periods of the oscillating signal~\cite{centers_stochastic_2021} and our measurement time $T$ is 12~hours. 
For $T\gg\tau_c$, the field is deterministic and the local dark-matter density averages to $\rho_{\text{\tiny{DM}}}=0.4$~GeV/cm$^3$~\cite{weber_determination_2010,catena_novel_2010}\footnote{This corresponds to a value of $\rho_{\text{\tiny{DM}}}=3.1 \times 10^{-42}$~GeV$^4$ in natural units. }. 
If $T\ll\tau_c$, the field is stochastic and the amplitude then follows a Rayleigh distribution with scale parameter 
$a_0/ \sqrt{2}$~\cite{foster_revealing_2018}. 
Since our measurement time and frequency range cover both cases, we display both limits in Fig.~\ref{fig:exclusionPlot}.
The upper limit at a given frequency is calculated by integrating the normalized distribution of the coupling $C_G / f_a$ up to the confidence limit of 95\%. Hence, the upper integration constant corresponds to the upper limit of the ALP-gluon-coupling. 
In the case of deterministic dark matter, the coupling follows solely a Rayleigh distribution. However, in the case of stochastic dark matter, the coupling corresponds to the ratio of two Rayleigh distributions which has a much longer tail, resulting in a higher upper limit. We determined a scaling factor of $3.2 \pm 0.3$ compared to the deterministic limit, in agreement with similar calculations by the CASPEr collaboration \cite{garcon_constraints_2019}. 
\\ 
Figure~\ref{fig:exclusionPlot} shows our exclusion region of the ALP-gluon-coupling as a function of mass or frequency. The most stringent constraint of $C_G / f_a m_a = 2.7 \times 10^{13}$~GeV$^{-2}$ (95\%~C.L.) for deterministic dark matter was set in the frequency range between 5~mHz and 5~Hz~\footnote{As an example: an oscillating magnetic field amplitude of $B_a = 10$~pT which is typical for this frequency range (see Fig.~\ref{fig:analysisSteps}d) translates with Eq.~(\ref{eq:faxionCorrelation}) and $E = 2 \times 35$~kV/cm into an EDM amplitude of $d_n = 8.6 \times 10^{-24} \, e \, \text{cm}$. With the use of Eq.~(\ref{eq:axionCoupling}) this can be translated into a ALP-gluon-coupling of $C_G / f_a = 1.4 \times 10^{-11}$~GeV$^{-1}$ for an ALP mass of $m_a = 10^{-15}$~eV. It follows $C_G / f_a m_a = 1.4 \times 10^{13}$~GeV$^{-2}$. }. 
For frequencies below 5~mHz, the upper limit increases due to magnetic gradient field drifts.
For frequencies above 5~Hz, the upper limit increases due to a decrease in sensitivity of the apparatus, as shown in Fig.~\ref{fig:analysisSteps}d. 
For reasons of legibility, we smoothed the limits in Fig.~\ref{fig:exclusionPlot} with a Savitzky-Golay filter~\cite{savitzky_smoothing_1964}. 
To provide context, the constraints on 95\%~C.L. from the other laboratory experiments are also presented~\footnote{Note, other experimental constraints are sometimes presented using the following relation: $g_a = (3.7 \pm 1.5) \times 10^{-3} \left( \frac{C_G}{f_a} \right) \frac{1}{\mathrm{GeV}}$~\cite{zyla_review_2020,ohare_cajohareaxionlimits_2020}.}.
In addition, indirect astrophysical and cosmological constraints arise from galaxy luminosity functions at high red-shifts, big bang nucleosynthesis (BBN) models, and the SN1978A cooling. The QCD-axion line shows the region where an axion would simultaneously solve the strong CP problem and explain all dark matter. \\


In conclusion, we performed a direct laboratory search for axion-like particles but did not find a significant oscillating signal. We could constrain an ALP-gluon-coupling in a mass region covering almost eight orders of magnitude. Together with the results of two other experiments, a large region of the ALP-dark-matter parameter space could be excluded, and future EDM searches may extend this even further. \\

\begin{acknowledgments}
We gratefully acknowledge the excellent technical support by R.~H\"anni, J.~Christen, L.~Meier, and D.~Berruyer.
We thank O.~Zimmer for lending us the 35~kV bipolar power supply, which was essential for data-taking.
This work was supported via the European Research Council under the ERC Grant Agreement no.\ 715031 (BEAM-EDM) and via the Swiss National Science Foundation under grants no.\ PP00P2-163663 and 200021-181996.
\end{acknowledgments}

\nocite{apsrev42Control}
\bibliography{references.bib}

\end{document}